\documentclass[manuscript]{acmart}

\AtBeginDocument{%
  \providecommand\BibTeX{{%
    \normalfont B\kern-0.5em{\scshape i\kern-0.25em b}\kern-0.8em\TeX}}}

\setcopyright{rightsretained}
\copyrightyear{2023}
\acmYear{2023}
\acmDOI{XXXXXXX.XXXXXXX}

\acmConference[NN]{NN}{NN}{2023}
\acmPrice{}
\acmISBN{}

\usepackage{todonotes}
\usepackage{subfig}
\usepackage{tabularx}
\usepackage{longtable}

\begin{document}

\title[Enhanced Caption Designs for Language Learning]{Useful but Distracting: Keyword Highlights and Time-Synchronization in Captions for Language Learning}

\author{Fiona Draxler}
\orcid{0000-0002-3112-6015}
\affiliation{%
  \institution{LMU Munich}
  \city{Munich}
  \country{Germany}
  \postcode{80539}
}

\author{Henrike Weingärtner}
\orcid{0000-0003-1100-312X}
\affiliation{%
  \institution{LMU Munich}
  \city{Munich}
  \country{Germany}
  \postcode{80539}
}
\email{henrike.weingaertner@ifi.lmu.de}

\author{Maximiliane Windl}
\orcid{0000-0002-9743-3819}
\affiliation{%
  \institution{LMU Munich}
  \city{Munich}
  \country{Germany}
  \postcode{80539}
}
\email{maximiliane.windl@ifi.lmu.de}

\author{Albrecht Schmidt}
\orcid{0000-0003-3890-1990}
\affiliation{%
  \institution{LMU Munich}
  \city{Munich}
  \country{Germany}
  \postcode{80539}
}
\email{albrecht.schmidt@ifi.lmu.de}

\author{Lewis L. Chuang}
\orcid{0000-0002-1975-5716}
\affiliation{%
  \institution{TU Chemnitz}
  \city{Chemnitz}
  \country{Germany}
  \postcode{09111}
}
\email{lewis.chuang@phil.tu-chemnitz.de}

\renewcommand{\shortauthors}{Draxler et al.}

\newcommand{\original}{\textit{Standard Captions}}
\newcommand{\highlighted}{\textit{Keyword Highlights}}
\newcommand{\karaoke}{\textit{Timed Keyword Highlights}}
\newcommand{\timed}{\textit{Timed Keywords}}

\def\subsectionautorefname{Section}

\begin{abstract}
Captions provide language learners with a scaffold for comprehension and vocabulary acquisition. Past work has proposed several enhancements such as keyword highlights for increased learning gains. However, little is known about learners' experience with enhanced captions, although this is critical for adoption in everyday life.
We conducted a survey and focus group to elicit learner preferences and requirements and implemented a processing pipeline for enhanced captions with keyword highlights, time-synchronized keyword highlights, and keyword captions. A subsequent online study ($n = 49$) showed that time-synchronized keyword highlights were the preferred design for learning but were perceived as too distracting to replace standard captions in everyday viewing scenarios. We conclude that keyword highlights and time-synchronization are suitable for integrating learning into an entertaining everyday-life activity, but the design should be optimized to provide a more seamless experience.
\end{abstract}

\begin{teaserfigure}
    \centering
    \includegraphics[width=\textwidth]{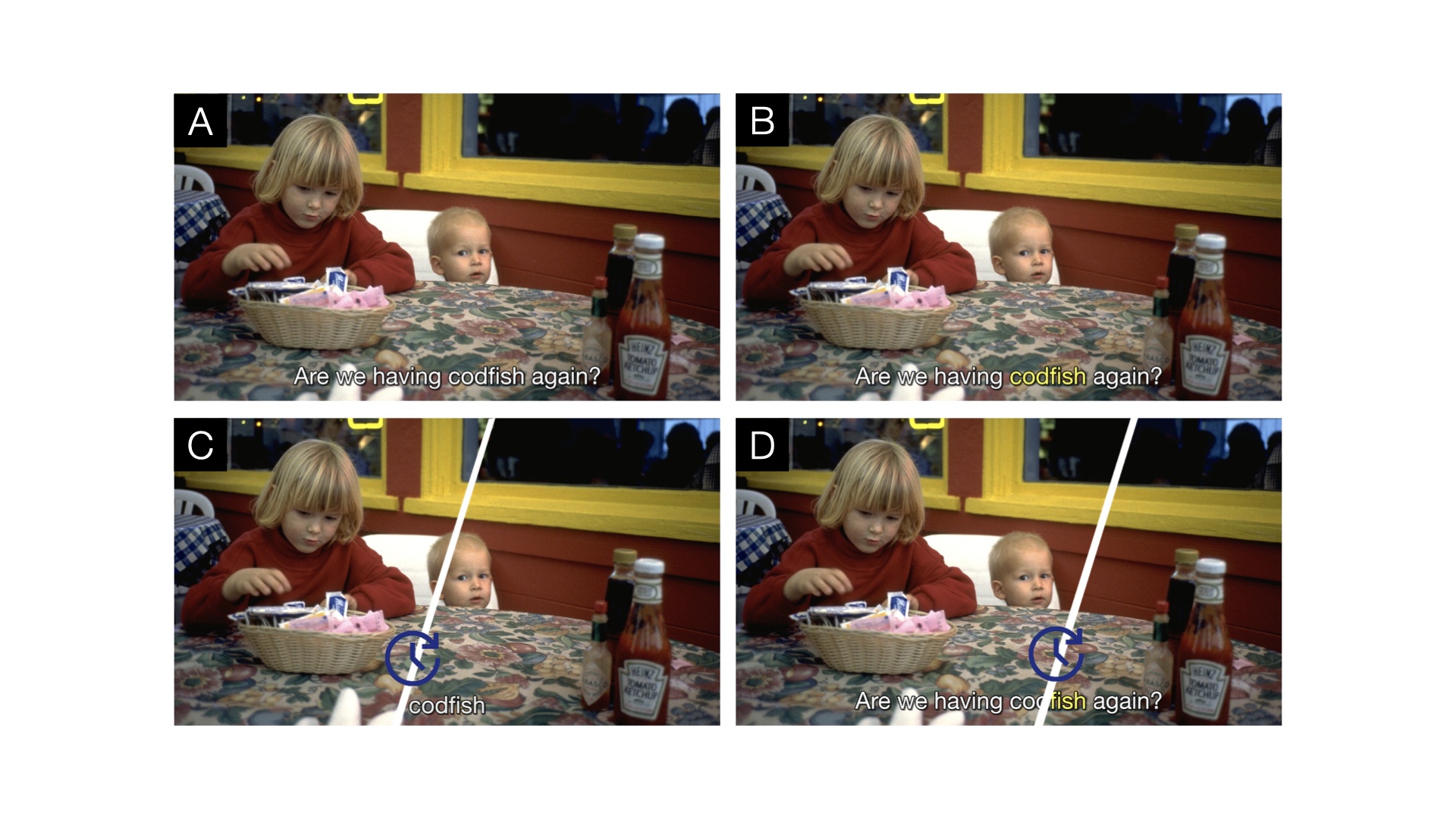}
    \caption{The four selected caption designs. (A) Standard captions, (B) full captions with keyword highlights, (C) timed keyword-only captions, (D), full captions with timed keyword highlights, where each keyword is highlighted when it is pronounced.}
    \label{fig:caption_design}
    \Description{Visualization of the four caption types on top of a movie still. (a) shows standard captions (white text with black contour), (b) adds yellow keyword highlights to the standard captions, (c) keyword-only captions where each keyword is shown at the moment it is pronounced, (d) identical to the keyword highlights, except that keywords are only highlighted when they are pronounced.}
\end{teaserfigure}



\maketitle

\section{Introduction}

With streaming services and online video platforms, language learners have gained access to potentially unlimited content. Thanks to foreign-language audio and captions, they can improve their skills while watching their favorite show. However, captions on streaming platforms and other media providers are primarily designed for comprehension, not for engaging learners. For example, they include potentially distracting elements such as textual sound descriptions (e.g., [footsteps approaching] or [\textit{Dancing Queen} playing on the radio]). Thus, optimizing captions to match language learners’ needs could improve the motivation to watch foreign-language media with captions and, in turn, also increase learning success.

Past work has already explored modifications of captions such as keyword captions~\cite{montero_perez_enhancing_2015}, captions including keyword translations~\cite{kovacs_smart_2014}, or interactive support based on eye tracking~\cite{fujii_subme_2019}. In fact, several studies show increased learning gains for such enhanced captions \cite{kovacs_smart_2014,cintronvalentin_captioning_2019,teng_vocabulary_2022}. However, the learners' perspective on enhanced captions is unclear, although a positive user experience may motivate viewers to integrate learning into leisure activities. %

In this paper, we applied a user-centered design process to implement and evaluate the user experience and perceived usefulness of enhanced closed captions for language learning, targeting medium- to high-proficiency learners. As a first step, we identified learner needs in a focus group and an initial survey. Based on related work and our insights from the survey and focus group, we implemented a processing system for three enhanced caption types: (1) captions consisting only of time-synchronized keywords, (2) captions with keyword highlights, and (3) captions with time-synchronized keyword highlights. Words were considered keywords if they were not included in an English-language CEFR\footnote{European Reference Scale; \url{https://www.coe.int/en/web/common-european-framework-reference-languages/level-descriptions}} A1-B1 corpus. As a baseline design, we added standard full captions. We compared the viewing experience and perceived understanding with these four caption types in an online survey using excerpts from the movie Marriage Story. We found that (time-synchronized) \highlighted{} captions outperformed \original{} and \timed{} questions with regards to hedonic qualities and scored almost as high as \original{} on pragmatic qualities and perceived comprehension. However, the distractions caused by the highlights meant that a majority of users still preferred standard captions, except when they explicitly aimed at learning.

In sum, we contribute (1) a choice of three enhanced caption types that are promising from a user perspective, (2) a comparative evaluation of these caption types with regard to user experience and perceived comprehension, and (3) a discussion of implications for embedding captioned viewing in everyday life to support language learning.

\section{Related Work}

Foreign-language videos, be it movies or TV shows, are a great tool for language learning: they immerse learners in a foreign culture~\cite{istanto_use_2009}, enable comprehension practice~\cite{nation_four_2007}, and promote vocabulary learning~\cite{rodgers_incidental_2020,peters_incidental_2018}. %
Generally speaking, videos provide exposure to authentic language, which is beneficial for language acquisition according to Krashen's input hypothesis~\cite{krashen_power_2004, eckman_teaching_2014}.
This section summarizes how learning can be supported through captions and subtitles\footnote{We use the term \textit{captions} to refer to intralingual or same-language subtitles and \textit{subtitles} to refer to interlingual or foreign-language subtitles~\cite[p. 9]{vanderplank_captioned_2016}}.
We discuss advanced caption design concepts that utilize the flexibility of current-day media players to optimize the viewing and learning experience for different target groups and briefly address technological prerequisites.

\subsection{Captions and Subtitles for Language Learning}

Captions and subtitles foster language learning through improved content comprehension~\cite{bianchi_captions_2008}, listening comprehension~\cite{guillory_effects_1998}, vocabulary acquisition~\cite{montero_perez_captioned_2013,hsu_effects_2012}, and to some extent, also grammar learning~\cite{cintronvalentin_captioning_2019}.
For example, a study on content and listening comprehension showed that students who watch videos with subtitles or captions write better summaries than students without captions~\cite{markham_effects_2001}. Similarly, learners provided with captions achieved higher scores in comprehension questions than those without~\cite{guillory_effects_1998}.
In terms of vocabulary learning, studies have observed both recall and recognition improvements when watching videos with captions or subtitles \cite{montero_perez_captioned_2013,hsu_effects_2012}. How many words a viewer learns depends on factors such as the words' imagery potential and whether the words sound similar to first-language words \cite{peters_effect_2019}. 
Interestingly, \citet{chen_transient_2018} found larger vocabulary gains for more proficient students. 
Studies on grammar learning through subtitles are scarce overall. For example, \citet{cintronvalentin_captioning_2019} found positive effects of textual enhancements in captions, but only for some of the enhanced structures, while a study with children by \citet{lommel_foreigngrammar_2006} showed no effects on grammar learning.
One important aspect to consider for learning success is cognitive load. On the one hand, the combination of multiple modalities---the associations of images, written, and spoken words---supports dual coding~\cite{danan_reversed_1992,mayer_using_2017} and can lead to a greater depth of processing~\cite{danan_captioning_2004}.
On the other hand, subtitles add an additional information channel that viewers need to process, and this can potentially cause a high cognitive load. Accordingly, a study by \citet{taylor_perceived_2005} showed that many first-year learners found captions distracting and that adding captions impacted their listening comprehension. However, this was not the case for third-year learners who already had more language exposure. Similarly, an eye tracking and EEG study by \citet{lacruz_chapter_2018} showed that despite the verbal redundancy effect, the risk of cognitive overload caused by captions was low.
Therefore, our target group in this work is also medium- to high-proficiency learners.
An outlook on additional aspects, such as the suitability of different video genres and recent work on learner strategies, is provided in the literature reviews by \citet{vanderplank_captioned_2016} and \citet{montero_perez_second_2022}.

The cited literature above includes work on captions (i.e., subtitles in the same language as the video) and subtitles (i.e., subtitles in the users' language). In fact, research so far has not shown conclusive evidence in favor of one or the other \cite{matielo_effects_2015}. Unsurprisingly, subtitles are particularly helpful for content comprehension of novice learners~\cite{bianchi_captions_2008}. However, another study found that learners watching a video with Scottish or Australian accents and English captions were better at understanding and repeating words than a Dutch subtitle group~\cite{mitterer_foreign_2009}. Regarding vocabulary learning, a 7-week study by \citet{frumuselu_television_2015} indicated that both novice and advanced learners perform better when using captions.
Moreover, \citet{markham_effects_2001} suggest advancing from subtitles to captions to no captions on subsequent viewings as a beneficial strategy. 

In sum, the decision to use captions or subtitles depends on the learner's goal and context.
In this work, we focus on intralingual captions because of their widespread availability, or as \citet{vanderplank_captioned_2016} put it:

\begin{quote}
    \textit{We are [...] fortunate that those with a disability have provided us, who are merely `hard-of-listening' in a foreign language, with a wonderful resource not only for making films and TV programmes accessible to us but for helping us improve our reading, listening, and speaking skills.}
\end{quote}

\subsection{Enhanced and Interactive Caption Design}

Above, we discussed standard full-text subtitles and captions. However, with current-day media players, loading new subtitle files has become very easy. This opens up new possibilities for static, adaptive, or even interactive subtitle and caption variants.
For example, static subtitle adaptations include captions that only show keywords~\cite{hsu_effects_2012,guillory_effects_1998} or highlight target word~\cite{ma_interactivesubtitle_2018}. Both of these approaches can benefit learning by increasing the focus on target words or reducing distractions. Other proposed methods add keyword translations, similar to text glosses \cite{sakunkoo2009gliflix,teng_vocabulary_2022}.
However, a major challenge with keyword or highlight captions is the selection of appropriate keywords, as it is difficult to assess what learners already know. A common approach is to select words based on their frequency in corpora, such as the BNC/COCA lists for English \cite{nation_making_2016}. \citet{guillory_effects_1998} had experts choose the words that were deemed most difficult.
As a further adaptation, \citet{kurzhals_close_2017} proposed speaker-following subtitles, which clearly mark the connection between speaker and dialog content and, thus, may reduce eye strain by reducing saccade length~\cite{fox_integrated_2016, kurzhals_close_2017}. However, this approach requires advanced preprocessing. \citet{wang2022incidental} investigated the effectiveness of bilingual subtitles compared to captions, subtitles, and no subtitles using an eye-tracking study. Thereby they found that while bilingual captions lead to a higher meaning recognition, they can also be distracting as users tend to spend more time reading the translations than the new words in the target language. \citet{mirzaei_partial_2014} investigated synchronizing speech signal and keyword captions and found short-term enhancements on subsequent viewing of non-captioned videos in comparison to full captions and no captions.
Finally, several projects and studies have explored interactive subtitles. For example, \citet{kovacs_smart_2014} enhanced captions with features for interactive vocabulary lookup, line translation, video navigation, and transcription to an alphabet familiar to the learner. This increased vocabulary learning in comparison to dual-language subtitles. However, the information-dense subtitles led to viewing times between 10 and 12 minutes for 5-minute videos, thus substantially changing the experience from linear viewing. In addition, \citet{zhu_vivo_2017} designed a dictionary where entries are enriched with captioned video clips, including target word highlights and translations, resulting in higher vocabulary retention than with a traditional dictionary.
Commercial platforms such as FluentU\footnote{\url{https://www.fluentu.com}, last accessed 2023-01-20}, LingoPie\footnote{\url{https://lingopie.com}, last accessed 2023-01-20}, and Language Reactor\footnote{\url{https://www.languagereactor.com}} also provide interactive captions for language learning and promote this as an enjoyable way of learning.
Since our objective is to integrate learning using captions into everyday viewing experiences, we do not include interactive elements that may shift the focus toward learning and consequently impact entertainment and long-term motivation. Thus, we apply a static approach with preprocessed subtitle files.

\subsection{Subtitle Files and Subtitle Processing}

Srt files are well-suited for simple adaptations because they are human-readable and supported by common media players such as VLC and can even be activated on top of browser-based Netflix and other video-on-demand players with extensions such as Substital\footnote{\url{https://chrome.google.com/webstore/detail/substital-add-subtitles-t/kkkbiiikppgjdiebcabomlbidfodipjg}, last accessed 2022-09-05}.
However, they also come with several drawbacks. Notably, srt files are often unofficially distributed, are more easily available for blockbusters than arthouse movies, and frequently contain mistakes. In addition, the ideal timing can differ depending on the associated media type. For example, there may be additional opening credits in a BluRay version that are not shown by a video-on-demand provider, and this delays the timing of the BluRay subtitles, requiring manual synchronization or a tool such as~\cite{laiola_guimaraes_lightweight_2016}.

\section{Survey on Caption Usage}

As the first pointer towards favored caption designs for language learners, we surveyed 61 people on their current caption preferences and usage habits.
Specifically, we asked them how often they use captions or subtitles, what languages they set them to, and how much they like watching video material with captions. %

\subsection{Survey Participants}

The 61 respondents were recruited via university mailing lists. They were between 17 and 65 years old ($ M = 27.0$, $SD = 8.9$ years). Thirty-nine participants identified as female, 20 as male, one as diverse, and one did not disclose their gender. Fifty-nine participants were native German speakers, and two were native Russian speakers. Five participants listed a second native language (Italian, Russian, Farsi, or Spanish). %
The survey was conducted in German.
We incentivized participation with a raffle of 20€ vouchers (one per ten participants).

\subsection{Survey Results}

The survey results revealed diverse subtitling and caption habits and preferences.
This was already apparent from the caption usage within the last thirty days: 16\% of the participants reported never using captions, 26\% used them a few times per month, and 57\% used captions weekly or daily.
A majority of respondents stated that they used captions in the video language (74\%) or subtitles in their native language (45\%; multiple responses possible). 25\% also set subtitles to a third language, for example, when the available options are limited or when they are watching with someone else.
The primary reason for activating captions were insufficient language skills (74\%), distractions caused by a noisy environment (67\%), a low video volume (51\%), other people needing subtitles (51\%), a fast rate of speech (46\%), dialects (43\%), difficult words (38\%), for language learning (5\%), unintelligible pronunciation (3\%), or when watching without sound (3\%).
Responding to the phrase ``I like subtitles'', 46\% of participants agreed with the statement, 23\% reported a neutral feeling, and 31\% disagreed.

Overall, the survey highlights that subtitles and captions are frequently used. Most participants in our sample activate subtitles for better comprehension, whereas only a few intentionally do so for language learning.
This points to an opportunity to increase the motivation to learn by adapting the caption design.

\section{Focus Group on Preferred and Envisioned Caption Designs}

We conducted an online focus group with six participants to discuss how captions can be adapted to cater to the specific needs of language learners. First, we presented and discussed current caption solutions beyond traditional closed captioning. Then, we asked our participants to develop their own ideas. 

\subsection{Procedure and Participants}
The participants (three male, and three female) were between 20 and 30 years old. They were all native German speakers and had learned English in school.
After an introduction round, we showed the participants short video clips with caption designs from or inspired by prior work. We asked them to discuss the concepts in light of their usefulness for language learning.
The first five clips were shown in one go; the last three were presented one after the other whenever the conversation had come to a hold. Overall, we showed eight subtitle variants:
\begin{enumerate}
    \item Captions with translations and explanations for individual words as in \citet{zhu_vivo_2017}
    \item Captions with translations of words on hover as in \citet{kovacs_smart_2014}
    \item Captions with keyword highlighting and an additional text box with keywords and their translations as in \citet{ma_interactivesubtitle_2018}
    \item A modified version of the latter without highlights and translations
    \item Another modified version of \citet{ma_interactivesubtitle_2018} without the standard captions
    \item Captions with translations in parentheses as in \citet{sakunkoo2009gliflix}
    \item Displaying captions next to the person speaking as in \citet{kurzhals_close_2017}
    \item Rather than spoken words, the last variant presented in-place object labels and translations. This variant showcased caption use beyond dialogues.
\end{enumerate}

Following the discussion, the participants engaged in an ideation activity using the 6-3-5 brainwriting method\footnote{\url{https://en.wikipedia.org/wiki/6-3-5_Brainwriting}} on a collaborative board with digital sticky notes.
In the end, they shared and discussed their ideas with the group.
The focus group was conducted in German.

\subsection{Findings of the Focus Group}

The discussion in the focus group highlighted the importance of avoiding disruptions and considering cognitive demands while catering to situation-dependent information needs. The ideation phase provided a starting point for further exploration of adaptations and novel caption designs.

\paragraph{Disruptions and Cognitive Load}

The participants identified attention switches caused by the caption design as potential sources of disruption. They were also afraid that overloaded designs would make them miss parts of the movie. Our participants considered this particularly critical for caption variations that included translations and redundant or non-essential information.
Specifically, they emphasized that native-language translations immediately and automatically attract attention, limiting the resources available for the original captions and the scene content.
In addition, they found translations particularly distracting when the original caption and the translation used different alphabets.
When translations were to be displayed, participants preferred them to be positioned under the original word rather than in a separate keyword box to minimize lookup times.
Participants also said that only words that are actually pronounced should be displayed. Even for the use case of documentaries, they considered object label captions (keyword variant 8) not helpful because of the factual learning focus inherent to documentaries.
In sum, our participants were afraid they could not focus on more than one thing at a time.

\paragraph{Situation- and User-Dependent Information}

Participants noted that the requirements for captions depend on individual and situational factors such as the language level and the speakers' dialect or rate of speaking.
For example, they positively commented on the captions that moved along with speakers, in particular for speakers with strong accents or dialects. However, they felt that the display time might be too short for following fast speakers.
They also found the idea of keyword captions interesting. Keywords reduce the overall information load and can target words that are specifically helpful for learners of a given language level. For translated keyword captions, participants feared that they might not always be able to recognize them when they are pronounced.
Finally, they also discussed the timing of words so that they appear the moment they are pronounced. This way, viewers could immediately connect words with their pronunciation.

\paragraph{Extensions and Novel Ideas}

Based on the discussed caption designs and their own experience, the participants came up with novel ideas and extensions of the presented caption variants.
These ideas can be grouped into concepts that focus either on comprehension or learning.
For better comprehension, suggestions include selective captioning of characters that speak dialects or are hard to follow. Similarly, captions could highlight technical terms or words that occur particularly infrequently and are, thus, more likely to be unknown.
For learning, participants felt that it might be helpful to add or highlight homonyms, typical idioms, dialectal differences, and/or words without direct translations. In an interactive system, translations could be shown on request.
Grammatical support could be provided, e.g., by coloring different tenses, endings, word boundaries, or functions of words.
Moreover, the level of detail should be adaptable to match the viewers' language level.

\section{Final Caption Designs and Hypotheses}

The user-centered design process including the initial survey and focus group motivated our final selection of caption designs as detailed below. Before comparing the enhanced caption designs in a user study, we derive hypotheses regarding the expected effect on user experience, perceived comprehension, perceived learning, and vocabulary recall.

\subsection{Selected Caption Designs}

Based on past literature, the focus group, and the survey, we finally selected the following four caption designs that vary between focusing on target words through keywords and providing context through full captions (cf. \autoref{fig:caption_design}):

\begin{enumerate}
	\item \textbf{Standard full captions.} This variant represents state of the art and serves as a baseline.
	\item \textbf{Full captions with keyword highlights.} This variant is a modification of \citet{ma_interactivesubtitle_2018} that also proposed highlighting keywords. However, we do not show translations of the words because the participants in the focus group considered translations distracting. %
	\item \textbf{Timed keyword-only captions.} This caption type shows keywords at the exact time they are spoken, while all other words are removed. The idea is based on \citet{mirzaei_partial_2014}, who proposed timed keyword captions as a means to focus on vocabulary learning without the distraction caused by the full transcript. Our focus group also confirmed the potential of time synchronization.
	\item \textbf{Full captions with timed keyword highlights.} This variant is a hybrid of the timed keyword captions and the keyword highlights and was introduced to guide the viewers' attention while still providing context. With this variant, we aim to compensate for the potential mismatch between keyword selection and learner knowledge.
\end{enumerate}

\subsection{Hypotheses}
\label{sec:hypotheses}

We derive the following hypotheses concerning measures for user experience (UX), perceived comprehension and learning, and vocabulary recall. Assessing UX and perceived comprehension helps us understand what type of captions learners are potentially willing to use in everyday life. We also added vocabulary recall to position the effectiveness of our designs in relation to prior work, but this was not the primary focus for our study.
Hypotheses are based on related work, the focus group, and the survey. 

\begin{enumerate}
    \item[H1a:] The \textit{pragmatic quality} is rated highest for \original{}. We expect this as viewers know this variant and feel most comfortable using it.
    \item[H1b:] The \textit{hedonic quality} is rated highest for \karaoke{}. We expect this variant to be considered innovative and providing a good balance between context on the overall scene and focus on potentially challenging aspects. %
    \item[H2:] \karaoke{} and \highlighted{} achieve the best \textit{perceived comprehension}. Conversely, \timed{} achieve the lowest perceived comprehension. Again, we assume the focus on potentially challenging aspects to be crucial. Even though \citet{mirzaei_partial_2014} stressed the advantage of reducing captions to keywords and reducing reading times, we expect that the lack of context hinders understanding, especially when the keyword selection is not perfectly matched to the viewers' language level.
    \item[H3:] \karaoke{} and \highlighted{} fare best for \textit{perceived learning}. These are followed by \timed{} because viewers perceive a lack of context; \original{} are perceived as least suitable for learning.
    \item[H4:] Both highlighted variants and \timed{} improve \textit{vocabulary recall} scores in comparison to standard \original{}. As all three enhanced designs put additional focus on keywords, we expect them to attract the viewers' attention.
\end{enumerate}

\section{User Study}

To assess the hypotheses introduced in \autoref{sec:hypotheses}, we conducted a within-subject study with 49 participants.
Specifically, we compared the user experience, learning, and perceived comprehension with the four different caption types applied to four scenes from the movie \textit{Marriage Story}, a 2019 movie that follows a couple's divorce.
As one of the proposed top 10 movies for ``people at C1 level'' \cite{andrade_best_2020}, \textit{Marriage Story} is suitable for our target group of medium- to high-proficiency learners. The movie contains many dialogues, is non-violent overall, and it was easy to select non-explicit scenes with a diverse vocabulary.

\subsection{Caption Generation and Video Preparation}

\begin{figure}
	\centering
	\includegraphics[width=.95\textwidth]{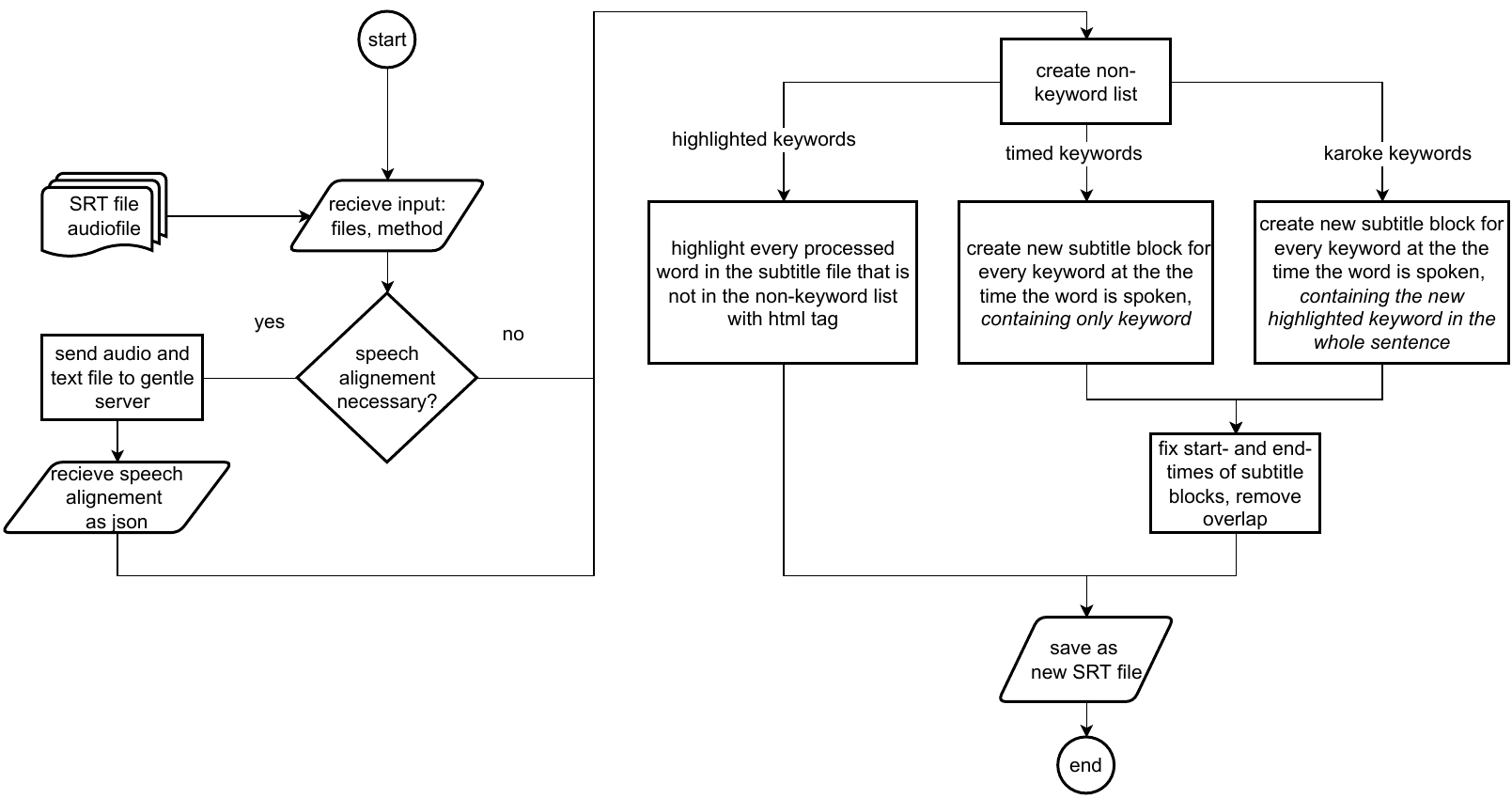}
	\caption{Processing pipeline for subtitle files}
	\label{fig:fig_pipeline}
	\Description{Flow diagram of our processing pipeline. Steps: (1) srt input, (2) if necessary: speech alignment with a gentle server, (3) creation of a non-keyword list, (4) condition-dependent manipulation of the srt files by removing non-keywords or highlighting keywords with HTML tags, (5) for timed conditions: processing overlaps, extending display times, (6) saving new srt files}
\end{figure}

We manipulate original srt files by removing non-keywords, adding highlights, or running forced alignment to adjust timestamps. \autoref{fig:fig_pipeline} gives an overview of the processing pipeline. We use a Python architecture with the pysrt package\footnote{\url{https://github.com/byroot/pysrt}} for working with subtitle files.
For all variants, the first step is the detection of keywords to determine what needs to be displayed or removed and what needs to be highlighted.
We follow a reverse approach, i.e., we mark a word as a keyword if it does not occur in non-keyword lists. For identifying words at a specific language level, we follow the approach proposed by \citet{andrade_best_2020}, who analyzed the vocabulary usage of a large number of movies. In particular, we merge the Oxford 5000 list\footnote{\url{https://www.oxfordlearnersdictionaries.com/about/wordlists/oxford3000-5000}} with the BNC/COCA corpus~\cite{nation_making_2016} to estimate the language level not only for the word stems but also the derived word forms and to remove proper names. When word levels are not uniquely identifiable (e.g., the stem ``accept'' is considered an A1 word, while ``acceptance'' is C1), we manually check for false positives. That is, we remove easy and frequent words that are not actually B2+ keywords.
We then mark keywords in the subtitles files with HTML font styling.
Finally, we run a Gentle\footnote{\url{https://github.com/lowerquality/gentle}} server for forced speech alignment. In case a keyword is highlighted for less than 500ms, we extend the display duration by 300ms or until the next caption line is shown.

We also used the script proposed by \citet{andrade_best_2020} to determine suitable scenes. 
For this, we evenly partitioned the subtitle file into 30 parts and counted B2+ word (keyword) occurrences in each part. We manually extracted scenes from high-keyword partitions and verified that the scenes did not include explicit content.
Finally, we prepared all four caption types for the resulting four movie clips of 2-3 minutes, leading to 16 preprocessed caption + video combinations. The video clips contained 24, 30, 39, and 41 keywords, respectively. Because of the higher density of keywords and partially overlapping speech, clips three and four were slightly more difficult than the first two.

\subsection{Procedure}

The study was implemented as an online survey and could be taken in Spanish or German. Once participants had read the study information and given their consent, we asked them about their experience with subtitles and their prior knowledge of English. We also included a vocabulary pre-test modeled after Nation's Vocabulary Size Test\footnote{\url{https://www.wgtn.ac.nz/lals/resources/paul-nations-resources/vocabulary-tests}, last accessed 2022-09-10}. The pre-test included multiple-choice questions on five keywords from each scene and four distractor items that did not occur in the videos. %
Participants then watched four movie clips, each with a different condition. Directly after each video, they responded to the UEQ-S~\cite{schrepp_design_2017} in the official Spanish or German version\footnote{Translations taken from \url{https://www.ueq-online.org}}. They rated their comprehension of the content and language and their overall impression of the caption variant.
The order of presentation and the pairing of the movie clip and caption variant were counterbalanced, using four of the 16 preprocessed videos for each participant.
After the four clips, we asked participants to what extent they had focused on learning, comprehension, and entertainment and asked them to rank the suitability of the caption variants for these goals. The last part of the survey was a vocabulary post-test.
Finally, two days later, participants took a second vocabulary post-test to accommodate for initial memory consolidation~\cite{dunlosky_improving_2013}.
We provide a full list of measures in \autoref{sec:study_measures}.
We collected the demographics via Prolific.

\subsection{Participants}

We recruited native Spanish and German speakers that did not live in English-speaking countries via Prolific\footnote{\url{https://prolific.co}}. Fourty-nine participants completed the study. Of these, 17 identified as female and 32 as male. They were between 19 and 49 years old ($ M = 30.3$, $ SD = 8.1 $ years). The 18 German speakers were residents of Germany (12), Austria (5), and Switzerland (1). The 31 Spanish speakers were residents of Mexico (16), Spain (9), Chile (5), and Portugal (1).
They self-assessed their English level at A2 (3), B1 (6), B2 (16), C1 (22), or C2 (6) on the CEFR scale.
The study took approximately 45 minutes, and participation was compensated with £8.5.

\section{Results}

This section presents the study results of with a focus on the participants' experiences and perceptions, following the hypotheses from \autoref{sec:hypotheses} and closing with a final ranking and outlook on participants' envisioned designs.

\subsection{Analysis}
We validate the hypotheses for the four caption types with a repeated-measures ANOVA, with \original{} serving as the baseline comparison. We apply a Greenhouse-Geisser correction when a Mauchly's test indicates a violation of the sphericity assumption. In case of a significant result, we follow up with pairwise post-hoc tests using a Holm correction and report Cohen's $ d $ for effect sizes.
We apply non-parametric Friedman tests with Holm-corrected Conover post-hoc tests for questions with a single ordinal scale.
All tests are performed with JASP~\cite{jasp_team_jasp_2022}.
To illustrate potential explanations of identified trends, we augment the report with exemplary participant statements\footnote{Translated to English if necessary}. For the preferred caption designs, we cluster all available responses and inductively derive general themes. %

\subsection{User Experience (H1)}

\begin{figure}
    \centering
    \includegraphics[width=.8\textwidth]{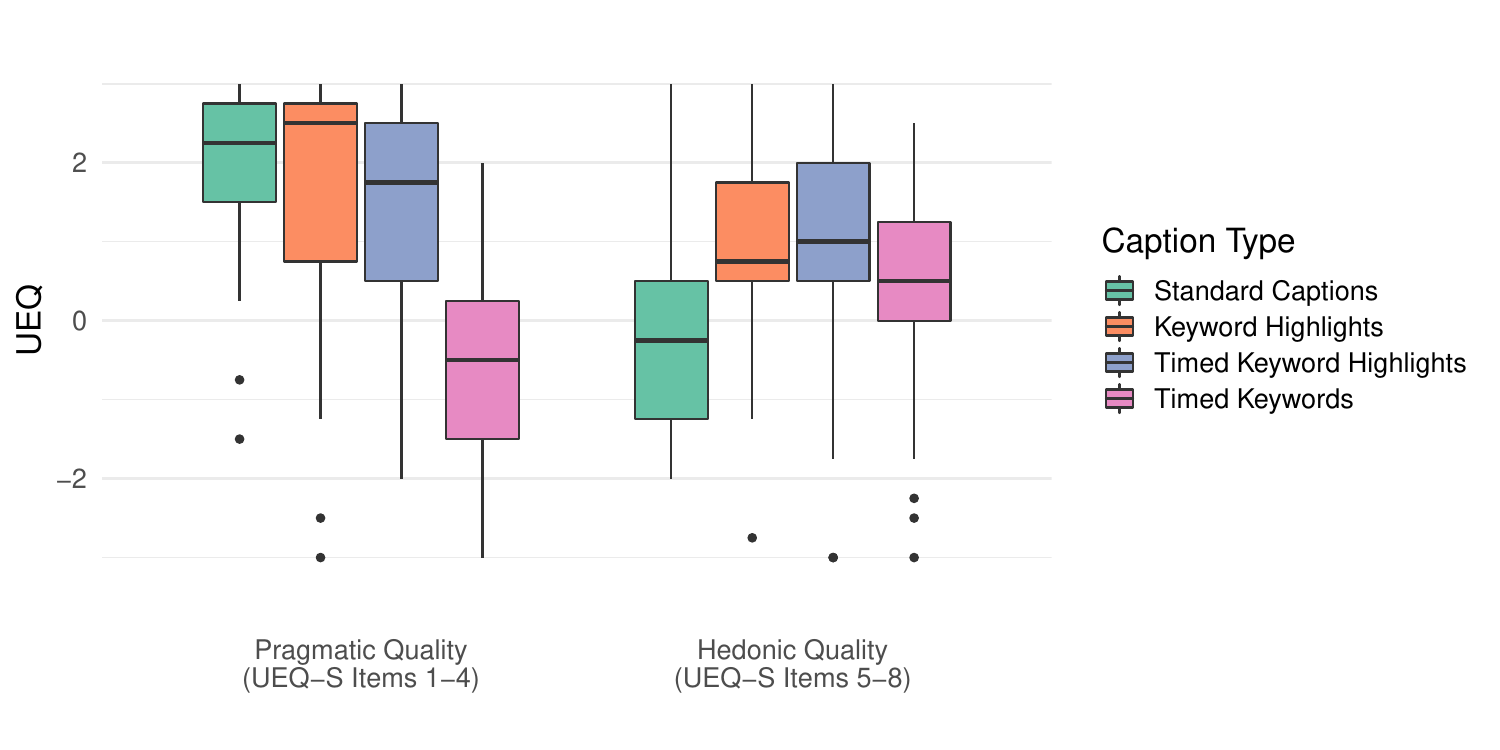}
    \caption{UEQ-S scores of the four caption types compared in our user study.}
    \label{fig:ueq}
    \Description{Boxplot of the UEQ-S scores, grouped by the pragmatic quality (UEQ-S items 1-4) and hedonic quality (UEQ-S items 5-8). The pragmatic quality is highest for keyword highlights, closely followed by standard captions and timed keyword highlights. Timed keywords perform significantly lower. The hedonic quality is lower overall. Keyword highlights and timed keyword highlights perform best, closely followed by timed keywords. Standard captions achieve the lowest average rating.}
\end{figure}

\begin{table*}[tb]
    \centering
    \caption{Median and standard deviation showing the agreement to opinion statements for each condition. Responses range from 1 (``I strongly disagree'') to 6 (``I strongly agree'').}
    \begin{tabularx}{\linewidth}{p{.29\textwidth} >{\centering\arraybackslash}p{.18\textwidth} *8{>{\centering\arraybackslash}X}}
        \toprule
        & & \multicolumn{2}{>{\centering\arraybackslash}p{.1\textwidth}}{\textbf{\original}} & \multicolumn{2}{>{\centering\arraybackslash}p{.1\textwidth}}{\textbf{\highlighted}} & \multicolumn{2}{>{\centering\arraybackslash}p{.1\textwidth}}{\textbf{\karaoke}} & \multicolumn{2}{>{\centering\arraybackslash}p{.1\textwidth}}{\textbf{\timed}}  \\ 
        \cmidrule(r){3-4} \cmidrule(r){5-6} \cmidrule(r){7-8} \cmidrule(r){9-10}
        & \textbf{\textit{Friedman test}} & \textbf{\textit{MD}} & \textbf{\textit{SD}} & \textbf{\textit{MD}} & \textbf{\textit{SD}} & \textbf{\textit{MD}} & \textbf{\textit{SD}} & \textbf{\textit{MD}} & \textbf{\textit{SD}} \\ 
        \midrule
        I understood the language well. & $ \chi^2(3) = 63.3 $, $ p < 0.001 $, $ W = 9.43 $ & 6 & 0.80 & 5 & 1.29 & 5 & 1.50 & 2 & 1.44 \\
        \midrule
        I understood the content well. & $ \chi^2(3) = 19.3 $, $ p < 0.001 $, $ W = 0.13 $ & 6 & 0.64 & 6 & 0.66 & 6 & 0.82 & 5 & 1.05 \\
        \midrule
        Viewing the video with this type of caption was agreeable. & $ \chi^2(3) = 35.0 $, $ p < 0.001 $, $ W = 0.24 $ & 4 & 1.34 & 5 & 1.10 & 5 & 1.47 & 3 & 1.54 \\
        \midrule
        I feel that I can learn new words very well with this caption variant. & $ \chi^2(3) = 23.0 $, $ p < 0.001 $, $ W = 0.16 $ & 6 & 0.55 & 6 & 0.47 & 6 & 0.76 & 5 & 1.03 \\
        \midrule
        I can very well imagine using this type of caption. & $ \chi^2(3) = 76.0 $, $ p < 0.001 $, $ W = 0.52 $ & 6 & 0.88 & 5 & 1.66 & 5 & 1.74 & 2 & 1.49 \\
        \bottomrule
    \end{tabularx}
    \label{tab:perception_statements}
\end{table*}

As seen in \autoref{fig:ueq}, \highlighted{} and \karaoke{} were rated best on the UEQ-S items representing the hedonic quality.
Pairwise post-hoc tests show significant differences between almost all conditions: \original{} fare worse than \karaoke{} ($ t = -5.17$, $ p < 0.001 $, $ d = -0.97 $), \highlighted{} ($ t = -4.86 $, $ p < 0.001 $, $ d = -0.91 $), and \timed{} ($ t = -2.56 $, $ p = 0.041 $, $ d = -0.48 $). \highlighted{} was rated better than \timed{} ($ t = 2.30 $, $ p = 0.046 $, $ d = 0.43 $), and so was \karaoke{} ($ t = 2.60 $, $ p = 0.041 $, $ d = 0.49 $).
With respect to the pragmatic quality, \timed{} were clearly outperformed by the three other conditions.
Accordingly, pairwise comparisons show that \timed{} performs significantly worse than \original{} ($ t = 10.29 $, $ p < 0.001 $, $ d = 1.96 $), \highlighted{} ($ t = 9.30 $, $ p < 0.001 $, $ d = 1.77$), and \karaoke{} ($ t = 8.10 $, $ p < 0.001$, $ d = 1.54 $). The remaining comparisons showed no significant differences.

In H1a, we posited that the pragmatic quality would be rated highest for \original. However, \highlighted{} and \karaoke{} performed similarly well.
As expected in H1b, the hedonic quality was highest for \karaoke{}, although \highlighted{} came close. Thus, the benefit of time-synchronization was not as large as expected.

\subsection{Perceived Comprehension of Language and Content (H2)}

As shown in \autoref{tab:perception_statements}, \original{} fared best for perceived language comprehension, closely followed by \highlighted{} and \karaoke{}. \timed{} achieved a very low overall score at $ MD = 2 $ and was significantly worse than all other conditions (all $ p < 0.01 $). All caption types substantially contributed to content comprehension, with no median score below 5 (out of 6).
This means that as predicted in H3, \timed{} achieved the lowest perceived comprehension. However, contrary to our expectations, \original{} was comparable for content comprehension and slightly better for language comprehension than \highlighted{} and \karaoke{}.

\subsection{Perceived Learning (H3)}

The high median of 5 or 6 for all caption types on the question ``I feel that I can learn new words very well with this caption variant,'' suggests that overall, participants considered all caption types helpful for learning (cf. \autoref{tab:perception_statements}). Conover post-hoc tests still indicated that \timed{} captions were significantly less suitable for learning than the other three types (all $ p \leq 0.01 $). %
There were no significant differences between the other conditions, and H3 cannot be confirmed.

\subsection{Vocabulary Recall (H4)}

The participants' prior knowledge of the tested vocabulary was high overall. On average, they correctly answered 88.0\% of the 24 questions in the vocabulary test before watching the videos, 87.5\% in the test right after, and 89.0\% in the 2-day delayed post-test.
There were no differences in the score changes from before watching the videos to the 2-day delayed post-tests when differentiated by caption type.
We observed clear ceiling effects: some participants already knew all the words tested for a condition and could, therefore, not improve their score.
In the survey, two people admitted that they looked up words, and several others may have done so.
All in all, we cannot confirm H4. We did not identify any differences in the keyword recognition scores.

\subsection{Final Ranking}

\begin{figure*}
    \centering
    \includegraphics[width=\textwidth]{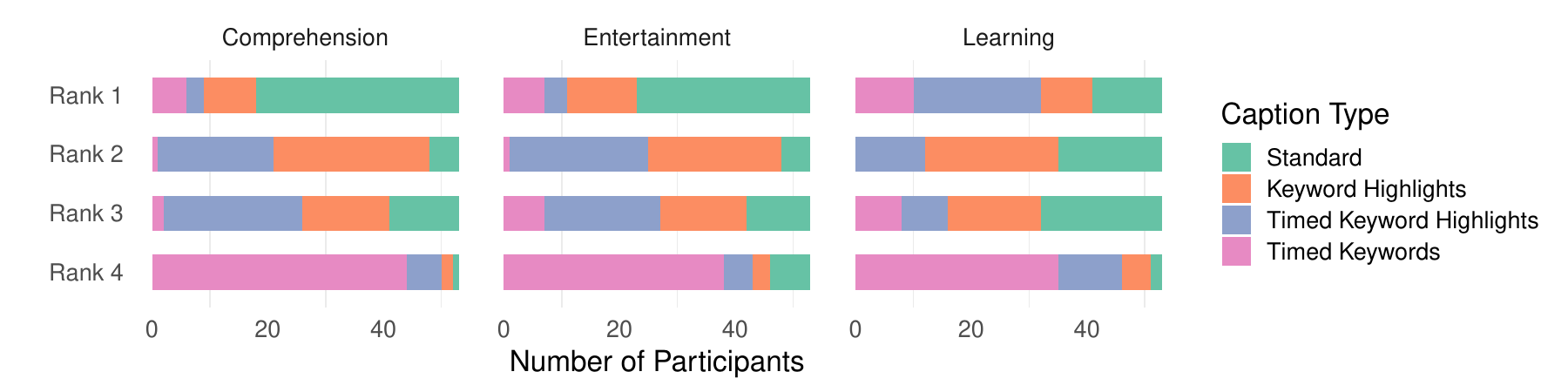}
    \caption{Ranking of the caption types for the purposes \textit{comprehension}, \textit{entertainment}, and \textit{learning}}
    \label{fig:rank}
    \Description{Standard captions are top-ranked for comprehension and entertainment, while timed keyword highlights are most frequently listed on Rank 1 for learning. Keyword highlights are most often mentioned on Rank 2. Timed keywords are the most frequently mentioned caption type on Rank 4 for all scenarios.}
\end{figure*}

The assessments above also align with the final ranking of the suitability for comprehension, entertainment, and learning after watching all videos (cf. \autoref{fig:rank}). \original{} captions were top-ranked for comprehension and entertainment, %
while \karaoke{} was top-ranked for learning. \timed{} obtained the lowest overall ranking for all three use cases.
This was also reflected in the absolute rating of the caption types: On a scale from 1 to 7, participants liked \original{} best ($ MD = 7 $, $ SD = 1.01 $). \highlighted{} ($ SD = 1.58 $) and \karaoke{} ($ SD = 1.72 $) were both rated at a median score of 6, and \timed{} at 3 ($ SD = 1.71 $). %

The participants' statements on the caption types give insights into possible reasons for the individual rankings.
Notably, \original{} captions were considered helpful for comprehension because they are \textit{``familiar''} (P49), \textit{``straightforward''} (P48), and \textit{``efficient and non-disruptive''} (P18). P12 described this type as \textit{``very clear, I understood everything perfectly.''} %
According to P21, they are \textit{``excellent for understanding spoken English in specific contexts.''} %
Typical comments explaining the participants' assessment of the \highlighted{} and \karaoke{} captions show that they were considered helpful but also distracting.
For example, for \highlighted, P27 noted that \textit{``as long as the video and audio are aligned, this type of viewing captions is agreeable to also learn sentence construction and figures of speech. Sometimes, it distracts from the video because it takes more time to read the full sentences.''} %
Similarly, P10 explained that \textit{``if you want to pay attention to [comprehension and learning], highlighted words distract a bit. I see their use when someone is trying to learn new vocabulary.''} %
P3 felt that \textit{``highlighting some words can make you loose time while reading because the brain will focus on this specific word.''} %
Time-synchronization tended to increase the perceived level of distraction:
P17 stated that they \textit{``started to think about which word will turn yellow next''} %
and P18 added that \textit{``The yellow words can be a bit distracting for people that already [know] pretty well the meaning.''}
Similarly, P7 liked seeing the highlights before they were spoken, so \textit{``you can anticipate the focus on the moment where it is mentioned.''} %
On the other hand, P49 found that \karaoke{} captions seemed to \textit{``support you in paying more attention to the plot than with `normal' subtitles.''} %
The comments also illustrate why some participants felt that \timed{} captions were not ideal for content and word comprehension. For example, P1 noted that they felt \textit{``distracted''} because this caption type was \textit{``more focused on drawing the attention towards certain words than on helping with the plot.''} %
Moreover, eleven participants explicitly mentioned that they lacked context when they only saw keywords or preferred types that provided full context.  %
For example, P42 said \textit{``The keywords alone do not contribute at all to the understanding of the context for me.''} %
Similarly, five participants found that showing all words was helpful for comprehension. %
Another issues was the selection of keywords: P27 noted that \textit{``the selected words did not necessarily coincide with [their] interest''}  %
and P42 found the highlighting of words in background conversations confusing. %

\subsection{Preferred Caption Designs}

\begin{table*}
    \centering
    \caption{Clusters of responses to the question ``If you could design your own captions, how would they look?'', including exemplary statements}
    \begin{tabular}{p{.9\textwidth}}
        \toprule
        \textbf{Additional elements -- Marking speakers}: 4 participants \\
        P10: \textit{I would assign colors to the characters so they can be distinguished more easily when several voices overlap} \\
        \midrule
        \textbf{Additional elements --- Translations, explanations, synonyms}: 9 participants  \\
        P19: \textit{They would be very similar to the timed keyword highlights, maybe with a synonym in brackets are including the translation of the word [...]} \\ %
        P21: \textit{With color codes that indicate if the highlighted words are verbs, nouns, etc.} \\ %
        P30: \textit{[...] with other words that are easier to understand and that are synonyms of the [keywords]} \\ %
        \midrule
        \textbf{Style suggestions, e.g., fonts}: 9 participants \\
        P1: \textit{I would focus on the clarity of the subtitles above all} \\ %
        P34: \textit{Simple, either white or yellow with a black contour so the font remains legible on a white background} \\ %
        \midrule
        \textbf{Keyword highlights (with minor changes)}: 15 participants \\
        P39: \textit{They would be a combination of standard subtitles with highlighted words in brackets} \\ %
        P45: \textit{maybe putting [keywords] a bit bigger than the other words or with a frame to mark the importance of the word} \\ %
        \midrule
        \textbf{Standard captions (with minor changes)}: 16 participants \\
        P14: \textit{I would simply leave it at the standard because it does a really good job and everyone is used to it} \\ %
        P41: \textit{Traditional captions because not everyone does not know the same words} \\ %
        P48: \textit{The truth is that standard captions are pretty similar to those I would design for my use} \\ %
        \bottomrule
    \end{tabular}
    \label{tab:preferred_captions}
\end{table*}

As an outlook, we asked participants how they would design their own captions. We clustered responses in \autoref{tab:preferred_captions}. Sixteen participants said they would stick to standard caption with no or almost no modification, largely because this is what they and other viewers are already used to.
Fifteen participants described a design very close to (time-synchronized) keyword highlights, adding some suggestions such as different typesetting. Thirteen participants listed additional elements to be included or changed in the captions, for example, different colors to distinguish speakers or background information on certain words.

\section{Discussion and Limitations}

By providing insights into the user perspective on captioned videos, we support researchers and practitioners in motivating users to embed learning activities into their everyday viewing experiences.
In particular, the opportunities and challenges we identified---such as the need for context, habits, distractions, and the potential to focus attention---inform the design of captioning for learning, comprehension, and entertainment.

\subsection{Distractions Outweigh the Potential of Enhanced Captions for Entertainment and Comprehension}

Although \highlighted{} and \karaoke{} performed better than or similar to \original{} on various measures, the overall ranking in \autoref{fig:rank} clearly shows that standard captions were the go-to solution in terms of comprehension and entertainment; only in the learning dimension, \karaoke{} overtook \original{}.
Specifically, \highlighted{} and \karaoke{} were similarly attractive alternatives on the pragmatic subscale of the User Experience Questionnaire and were rated higher on the hedonic subscale.
Similarly, the number of participants describing their preferred captions as a variant of \original{} or \textit{(Timed)} \highlighted{} captions was almost the same.
Still, it seems that due to the increased potential for distractions, the two caption variants that used highlighted keywords were not perceived as sufficiently agreeable, innovative, or helpful to overrule the influence of habits and familiarity. %
The ranking and participant statements further indicate that learners are only willing to accept divided foci of attention in a learning scenario.

Research on visual perception agrees that sudden and easily distinguishable stimuli attract a viewer's attention~\cite{corbetta_control_2002}. Thus, it is unsurprising that a colored and/or suddenly appearing keyword will achieve this.
So, while \citet{mirzaei_partial_2014} recommended timed keyword captions as a good alternative to standard captions because of the high density of relevant words, our findings suggest that participants did not like the viewing experience with timed changes and bright colors.

\subsection{Choosing Ideal Keywords is Hard -- Optimize Designs for Heuristics and Curricula}

We chose our keywords based on a word frequency corpus aligned with estimated language levels. This is a typical approach in language learning and was, for example, also used by \citet{mirzaei_partial_2014}. In other projects, keywords were based on expert ratings~\cite{guillory_effects_1998} or a pre-test~\cite{montero_perez_enhancing_2015}.
However, especially in our interconnected world and for a ubiquitous language such as English, it is almost impossible to perfectly model a learner's prior knowledge to predict unknown vocabulary. %
In fact, several participants in our study mentioned that the selected keywords did not match their expectations.
Moreover, watching movies is often a social experience including two or more people, and adding another person to the equation complicates the process even further.

This means that keyword highlights will, at most, be an educated guess. But how critical is this, really? We argue that a suitable caption design that balances distractions, context, and focus is more crucial.
In particular, we expect that highlighting a few words too many will not have a dramatic impact on the viewing experience, as long as they do not annoy or distract the viewer (as was the case in our study).
Consequently, we recommend a conservative selection of keywords. For example, less obtrusive examples, such as bold or italic print, could be used (see also \textit{textual enhancement} strategies~\cite{labrozzi_effects_2016}).
Furthermore, in the movie analyses performed by \citet{andrade_best_2020}, a substantial share of the vocabulary was estimated at B2 level or lower, indicating that the number of keywords in most movies will not surpass a certain threshold.
To preserve the context, the participants of our study demanded full captions. This is also beneficial with respect to imprecise keyword selection: full captions ensure that false negative keywords (unknown words that are not highlighted) will still be visible, albeit not highlighted.

Alternatively, captioned viewing could be aligned with classroom learning. We suggest a crowdsourced approach to collect target word lists. For example, \citet{culbertson_have_2017} proposed a system for correcting auto-generated captions that could be extended with a feature for learners to highlight words relevant to their language class.

\subsection{Limitations and Future Work}

Our initial hope was that our caption enhancements would foster learning without causing a negative impact on the viewing experience. If this were the case, there would be no reason for viewers to stick with standard captions.
However, enhanced captions were only top-ranked for a learning scenario. This highlights the need for further adaptations to make the viewing experience with enhanced captions similarly enjoyable. Currently, we do not know to what extent this preference was caused by our design choices, such as using the yellow color for highlights. Consequently, future work should analyze the effect of design choices, factoring in findings from label design~\cite{kruijff_influence_2019}.
We also encountered technical and methodological challenges during the implementation and evaluation of the caption types.
Notably, our processing pipeline is not yet fully automated and can, therefore, not be applied at scale.
For example, in two of the scenes we used, the lines of two characters partially overlapped. This required swapping some lines for the forced alignment, which our system is currently not capable of doing automatically.
In addition, although we aim to support implicit learning in everyday life, the constraints of our user study meant that we were not able to capture implicit learning directly. A long-term, in-situ study would be necessary to measure changes in the overall language level.

\section{Summary and Conclusion}

In this paper, we implement and evaluate three enhanced caption types that increase the focus on target words in language learning by highlighting and/or displaying words synchronized with the audio track. To gather viewers' opinions on these captions, we conducted an online survey evaluating the user experience, perceived comprehension, and vocabulary recognition with our enhanced caption types compared to standard captions. We discovered that participants preferred captions with highlights in a learning scenario but felt that they were too distracting for an everyday viewing experience. %
These findings highlight challenges in the widespread adoption of captions optimized for learning in language learners' everyday lives.

\bibliographystyle{ACM-Reference-Format}
\bibliography{sample-base}

\appendix

\section{Survey Measures}

\begin{longtable}{|p{.4\linewidth}|p{.2\linewidth}|}
\caption{Questions on subtitles and captions included in the online survey}
\label{tab:survey_measures}\\
\hline
\textbf{Question (translated to English)} &
  \textbf{Type of Question} \\ \hline
\endfirsthead
\endhead
I like to use subtitles/captions very much (any language). &
  5-point Likert scale \\ \hline
How often have you used subtitles/captions in the past 30 days? &
  Selection menu \\ \hline
In what situations do you use subtitles/captions? (any language)? &
  Selection menu with option to specify own \\ \hline
How do you set subtitles/captions when the video is in a foreign language (any language)? &
  Selection menu with option to specify own \\ \hline
If you could design your own subtitles/captions, how would they look? &
  Text field \\ \hline

\end{longtable}

\section{User Study Measures}
\label{sec:study_measures}

\begin{longtable}{|l|p{.4\linewidth}|p{.2\linewidth}|}
\caption{Measures and questions included in the user study}
\label{tab:study_measures}\\
\hline
\textbf{Measure} &
  \textbf{Question (translated to English)} &
  \textbf{Type of Question} \\ \hline
\endfirsthead
\endhead
Demographics &
  How old are you? &
  Text Field \\ \hline
Demographics &
  How do you identify yourself? &
  Selection menu with option to specify own \\ \hline
Demographics &
  In which country do you currently live? &
  Selection menu with option to specify own \\ \hline
Demographics &
  What level of education do you have? &
  Selection menu with option to specify own \\ \hline
Demographics &
  What is your current occupation? &
  Selection menu with option to specify own \\ \hline
Demographics &
  What is your native language? &
  Selection menu with option to specify own \\ \hline
English Experience &
  How often do you speak English? &
  Selection menu \\ \hline
English Experience &
  How often do you need to understand English (for example, when reading or on the Internet)? &
  Selection menu \\ \hline
English Experience &
  What is your English language level? &
  Selection menu \\ \hline
Vocabulary Pre-Test &
  What synonym or definition can you use to meaningfully replace the words in angle brackets in the following sentences? &
  4 Options per question \\ \hline
Caption Habits \& Preferences &
  I like to use subtitles very much (no matter in which language). &
  7-Point Likert Scale \\ \hline
Caption Habits \& Preferences &
  How often have you used subtitles (in any language) in the last 30 days? &
  Selection menu \\ \hline
Caption Habits \& Preferences &
  How do you set the subtitles if the video is in a foreign language (any language)? &
  Multiple Choice Selection menu with option to specify own \\ \hline
User Experience &
  UEQ-S &
  7-Point Likert Scale \\ \hline
Self-Assessment &
  I understood the language very well. &
  6-Point Likert Scale \\ \hline
Self-Assessment &
  I understood the plot very well. &
  6-Point Likert Scale \\ \hline
User Experience &
  Watching the video with this kind of subtitles was very pleasant. &
  6-Point Likert Scale \\ \hline
Self-Assessment &
  I have the impression that I can learn new words very well with this subtitle  variant. &
  6-Point Likert Scale \\ \hline
User Experience &
  I can very well imagine using this kind of subtitles myself. &
  6-Point Likert Scale \\ \hline
User Experience &
  I really like this subtitle variant overall. &
  7-Point Likert Scale \\ \hline
Additional Feedback &
  Is there anything else you would like to say? &
  Text field \\ \hline
Self-Assessment &
  How much did you pay attention to the following aspects while watching the videos? (Scene understanding, Learning new words, Entertainment) &
  6-Point Likert Scale for each \\ \hline
User Experience &
  Please sort all subtitle variants according to how well you like them if the focus is on learning new vocabulary. &
  Option to sort all 4 variants \\ \hline
Additional Feedback &
  Why did you sort the variants in this way? &
  Text field \\ \hline
User Experience &
  Please sort all subtitle variants according to how well you like them if the focus is  on entertainment/pleasure. &
  Option to sort all 4 variants \\ \hline
Additional Feedback &
  Why did you sort the variants in this way? &
  Text field \\ \hline
User Experience &
  Please sort all subtitle variants according to how well you like them when the focus is on scene comprehension. &
  Option to sort all 4 variants \\ \hline
Additional Feedback &
  Why did you sort the variants in this way? &
  Text field \\ \hline
Desired Captions &
  If you could design your own subtitles, what would they look like? &
  Text field \\ \hline
Vocabulary Retention &
  What synonym or definition can you use to meaningfully replace the words in angle brackets in the following sentences? &
  4 Options per question \\ \hline
Additional Feedback &
  Is there anything else you would like to say? &
  Text field \\ \hline
\end{longtable}

\end{document}